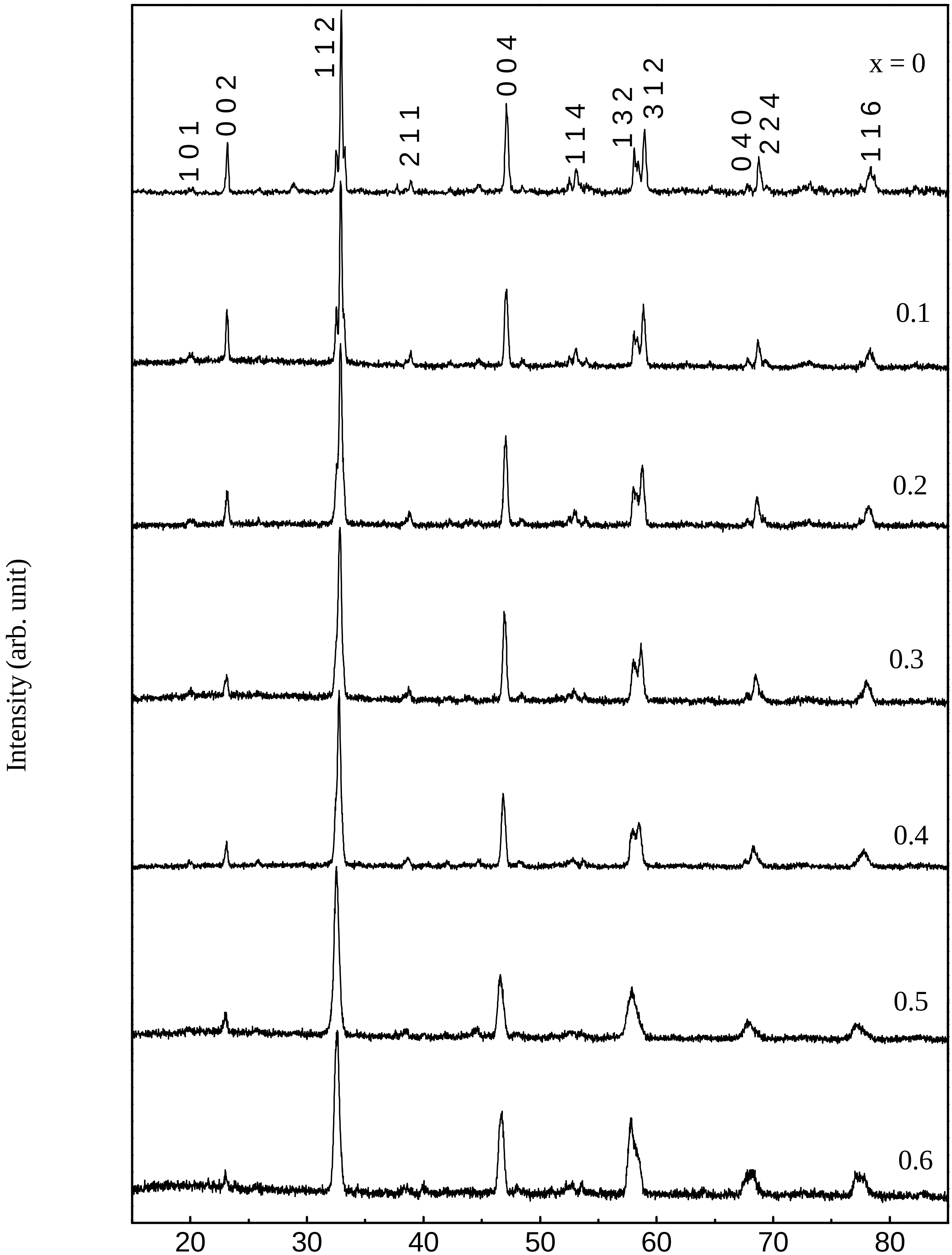

Fig.1 Room temperature XRD spectrum of $Ca_{2-x}La_xFeMoO_6$ samples

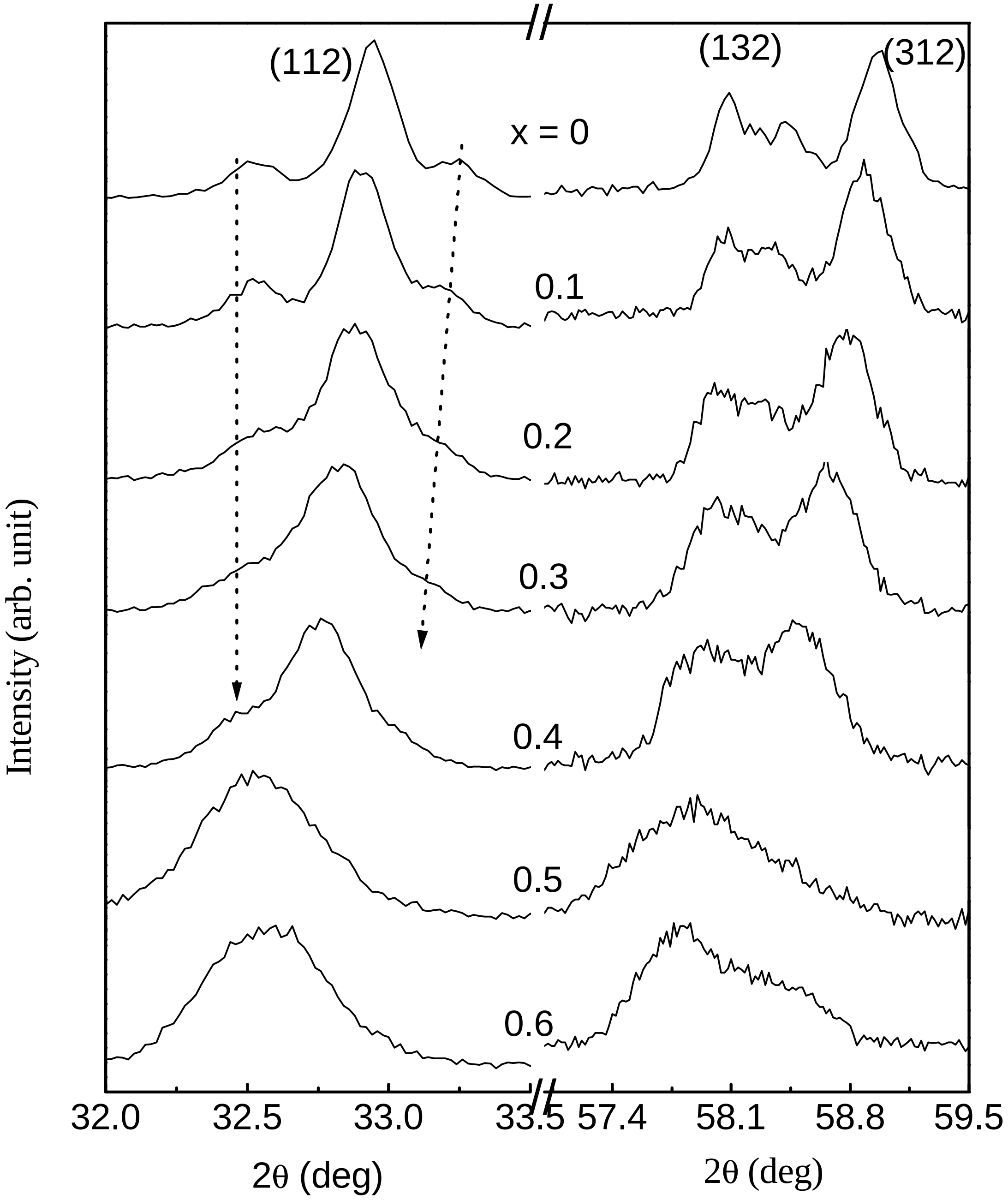

Fig.2 The spectral evolution at about (112), (132) and (312) are shown for $Ca_{2-x}La_xFeMoO_6$ samples. The dotted lines indicate the gradual decrease of wing type splitting for (112) peak.

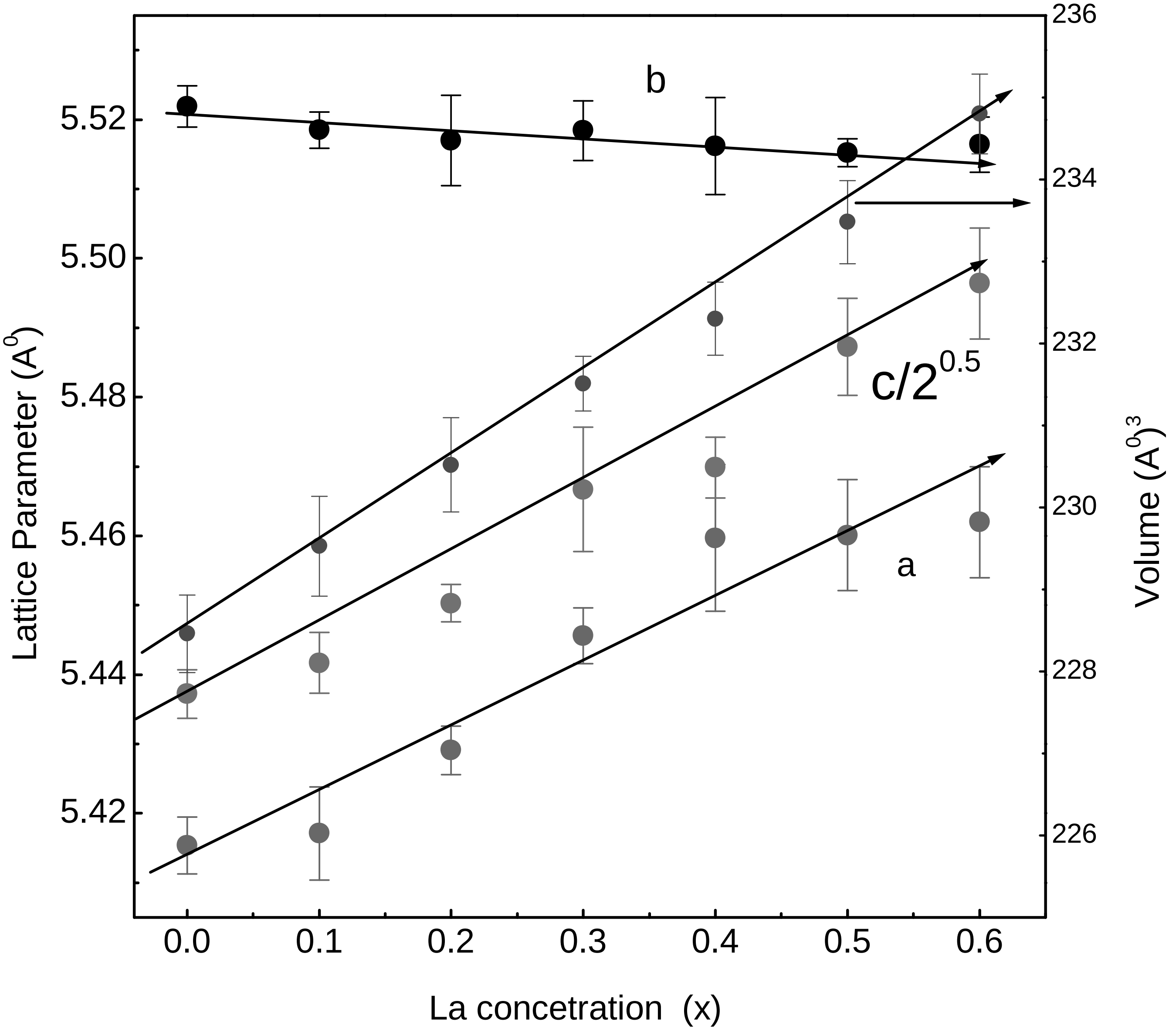

Fig. 3 (Colour online) Lattice parameters a, b, c and unit cell Volume with the variation of La concentration in $La_xCa_{2-x}FeMoO_6$ compound.

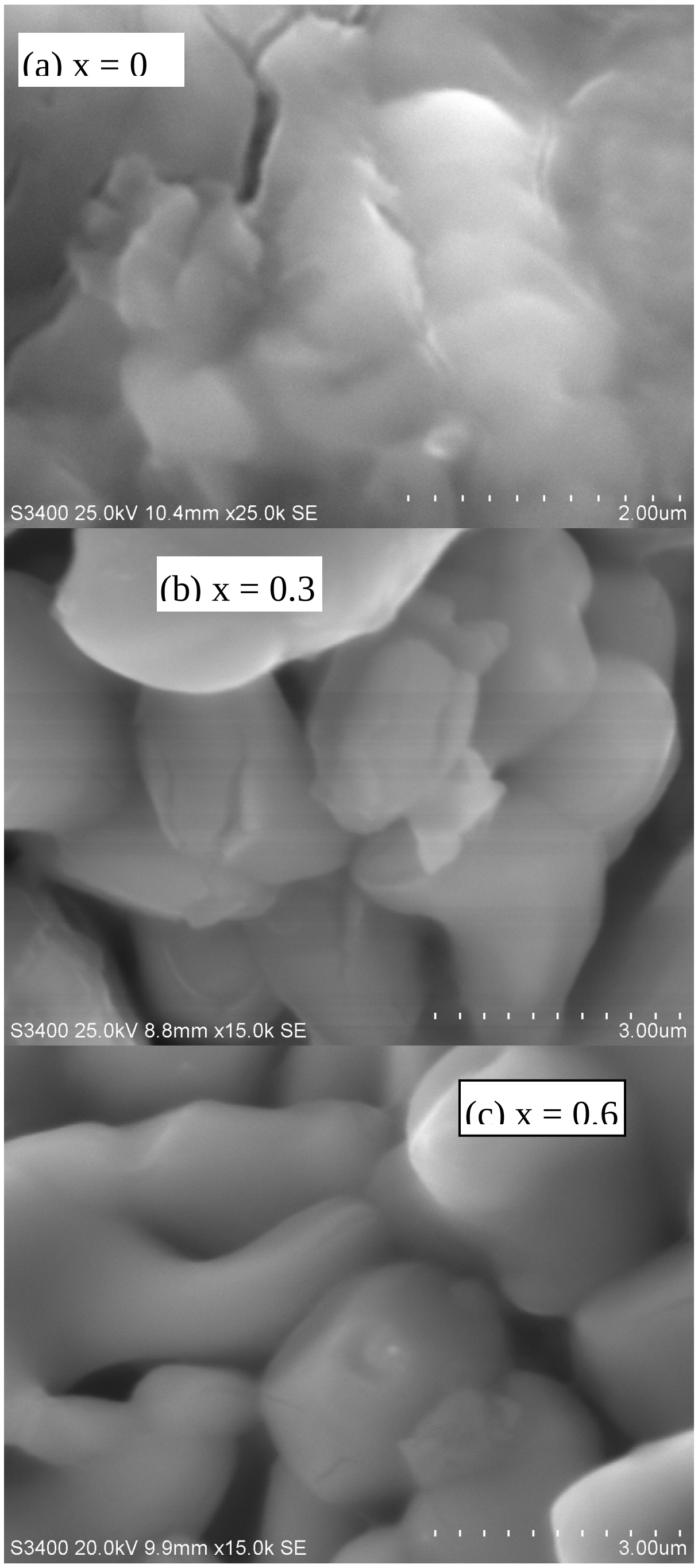

Fig.4 SEM picture of $Ca_{2-x}$ $La_xFeMoO_6$ compound in the 2-3 μm scale for La concentration x = 0 (a), x = 0.3 (b) and x = 0.6 (c) .

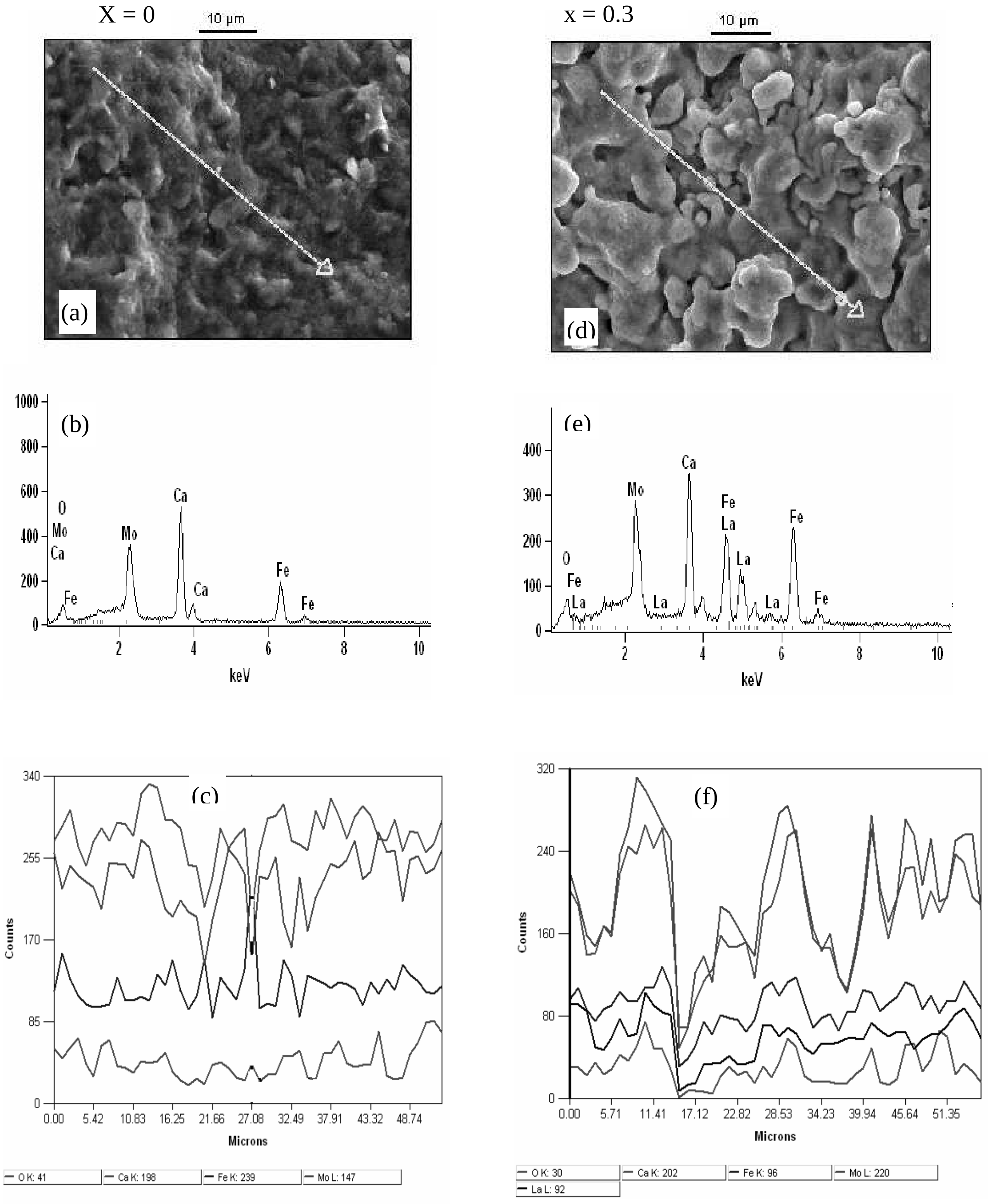

Fig.5 (Colour online) Experimental zone (a,d), EDX point spectrum (b, e) and line spectrum (c, f) for La concentration x = 0 (a-c) and x = 0.3 (d-f), respectively.

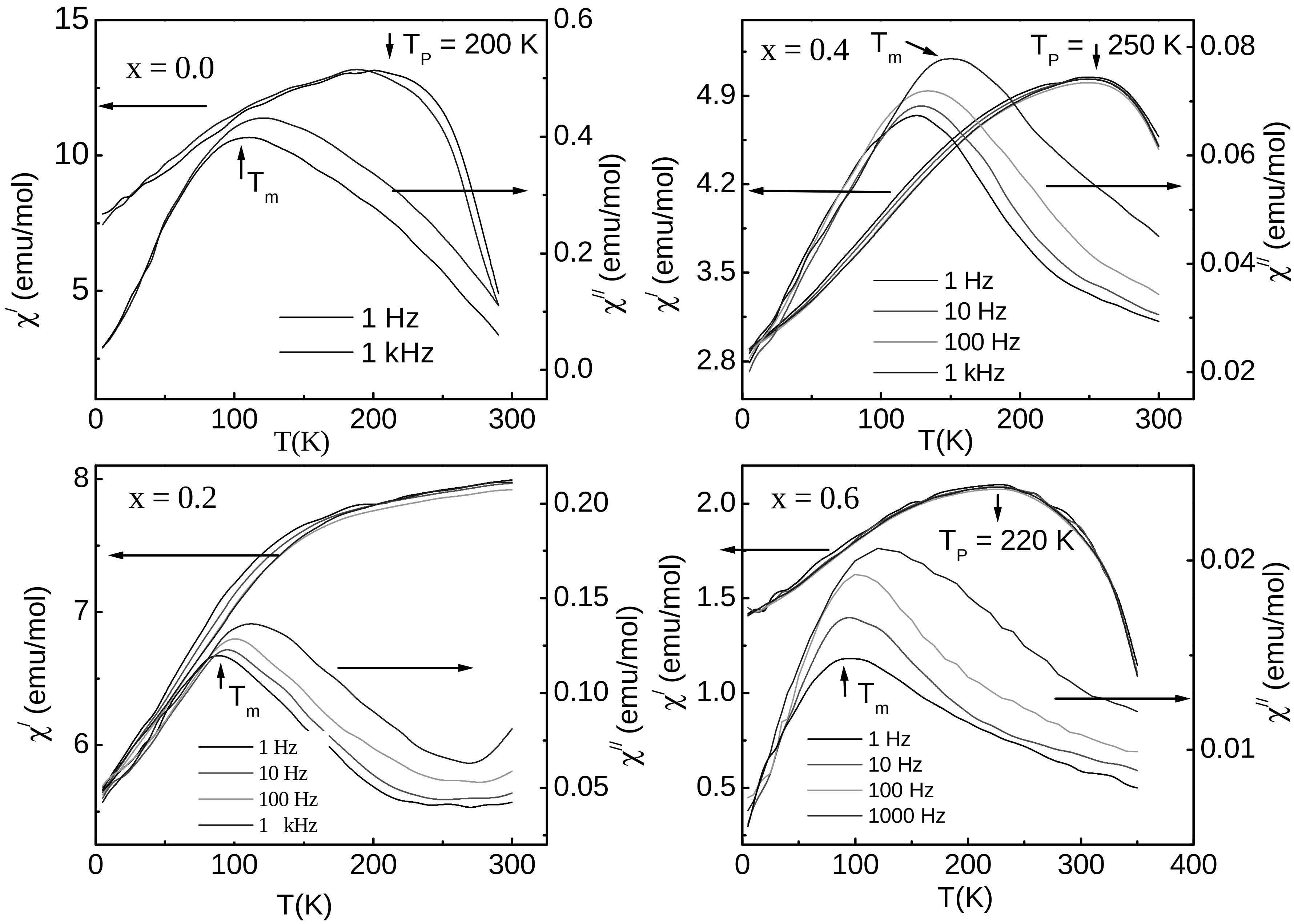

Fig.6 (Colour online) Temperature dependence of $\chi^{/}$ and $\chi^{//}$ for La concentration x = 0, 0.2, 0.4 and 0.6 samples.

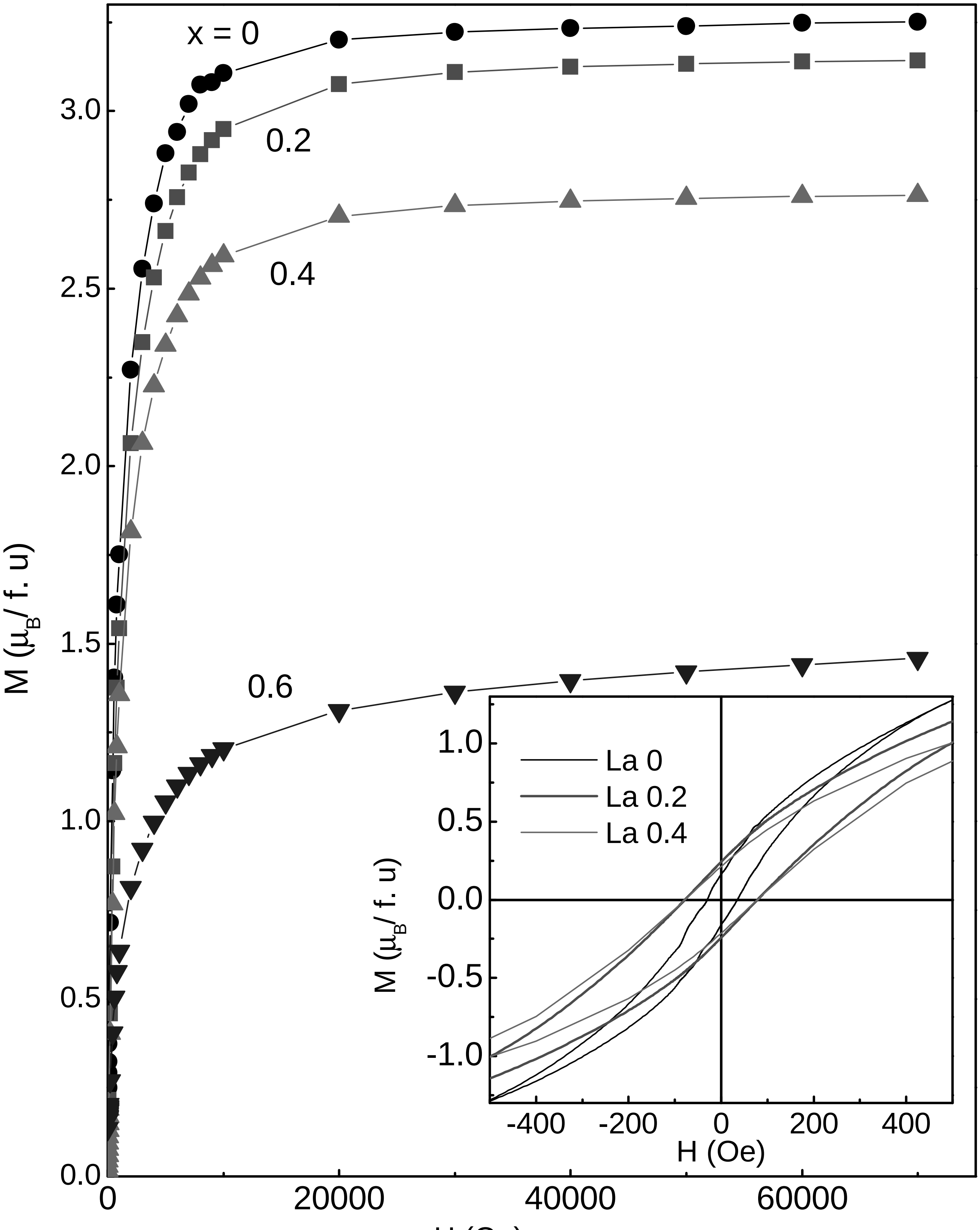

Fig. 7 (Colour online) Field (H) dependence of Magnetization (M) at 5 K for x = 0, 0.2, 0.4, 0.6 samples. The inset shows a typical loop for x = 0, 0.2 and 0.4 samples.

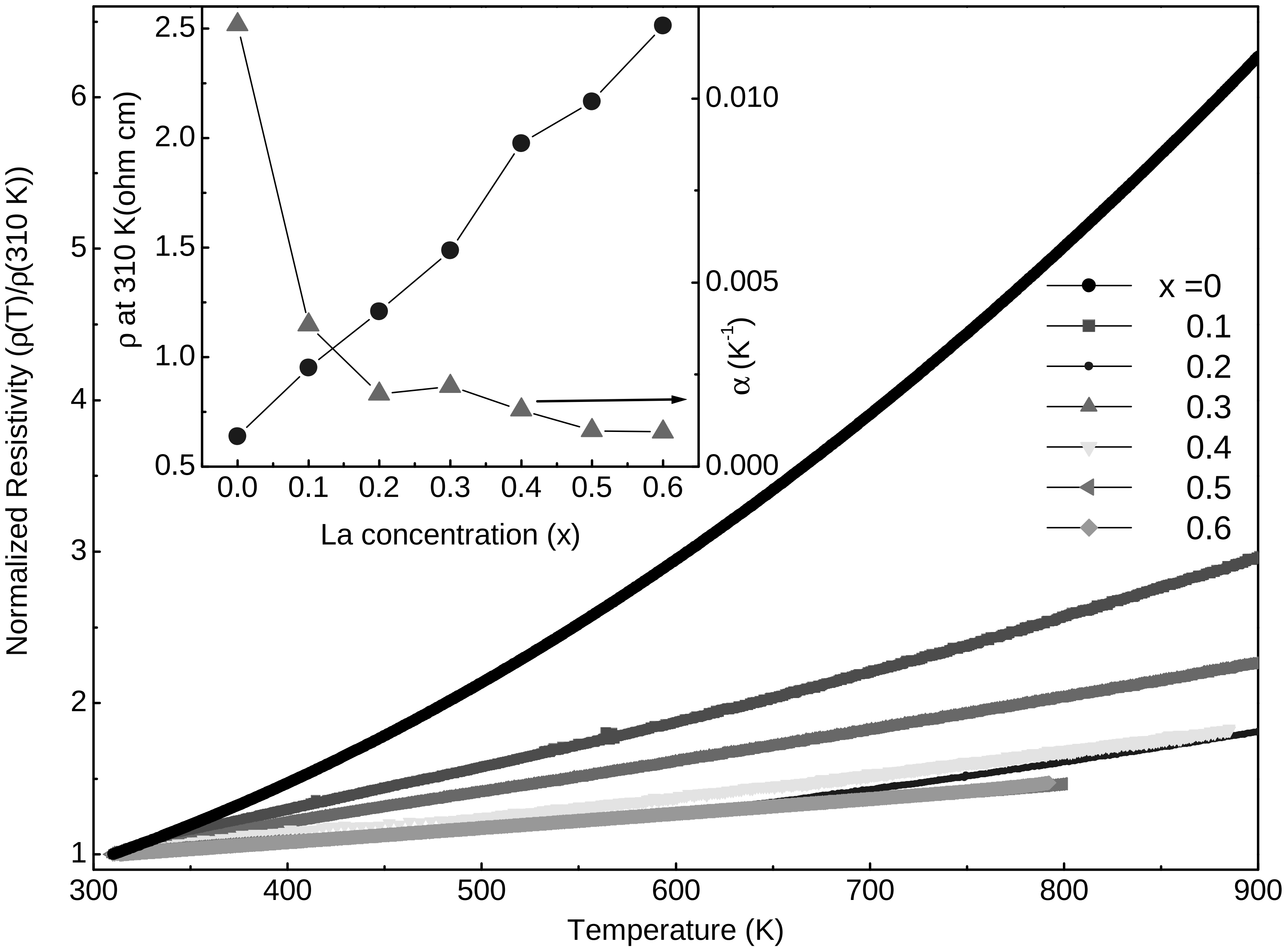

Fig. 8 (Colour online) Temperature dependence of resistivity (ρ) > 310 K. The resistivity data are normalized by 310 k data. The 310 K resistivity data and slope (α) of ρ (T) above 600 K are shown in the inset plot.

# Disorder Effects in La substituted ferrimagnetic $Ca_2FeMoO_6$ double perovskite

**I. Panneer Muthuselvam[1], Asok Poddar[2] and R.N. Bhowmik[1*]**

[1]Department of Physics, Pondicherry University, R. Venkataraman Nagar, Kalapet, Pondicherry-605014, India

[2]Experimental Condensed Matter Physics Division, Saha Institute of Nuclear Physics, 1/AF Bidhannagar, Kolkata-700064, India

[1*]Corresponding author: Tel.: +91-9944064547; Fax: +91-413-2655734

E-mail: rnbhowmik.phy@pondiuni.edu.in

Abstract

$Ca_{2-x}La_xFeMoO_6$ double perovskite with La concentration x = 0 to 0.6 was synthesized using solid state sintering route. The standard techniques of XRD, SEM and EDX were applied to characterize the material. Crystal structure of the samples was stabilized in monoclinic phase with space group $P2_I/n$ and lattice expansion was indicated with the increase of x. The increase of La concentration gradually suppressed the coexisting minor secondary phase in the material and simultaneously, EDX results indicated the accommodation of more Mo atoms in the crystal structure. A significant modification in the surface morphology of the material was noted from adhesive type surface for x = 0 to brittle type surface with more grain boundary contributions for La doped samples. We understand a significant change in magnetic properties (appearance of cluster glass component, reduction of magnetic moment and indication of higher $T_C$) and in electrical properties (reduction of metallic character) in terms of enhanced internal disorder in the material, introduced due to La doping in double perovskite structure.

I. INTRODUCTION

An ideal double perovskite structure is denoted by $A_2BB^{/}O_6$, where A is alkaline earth divalent cation ($Ca^{2+}$, $Sr^{2+}$, $Ba^{2+}$), B and $B^{/}$ are $Fe^{3+}$ and $Mo^{5+}$ transition metal ions, respectively. In this special structure of materials, two perovskite units, viz., $AFeO_3$ and $AMoO_3$, form alternatively rock-salt (NaCl) type arrangements with $FeO_6$ and $MoO_6$ octahedral units bonded by oxygen anions and A ions occupy the interstitial positions [1]. The localized spin model [2] predicted that $Fe^{3+}$ ($3d^5$, S = 5/2) induces antiferromagnetic ordering to $Mo^{5+}$($4d^1$, S=1/2) due to hybridization of 3d (Fe) and 4d (Mo) orbitals. The antiferromagnetic ordering of low spin $Mo^{5+}$ moments with respect to the high spin $Fe^{3+}$ moments results in the ferrimagnetic ground state in $A_2FeMoO_6$ compounds and electrically, the compounds are expected to be insulator. In fact, the materials may also exhibit the all possible properties of a ferromagnet. Experimental results confirmed the ferrimagnetic ground state [3], but the metallic behaviour in $A_2FeMoO_6$ below their $T_C$ is not consistent with localized spin model. The coexistence of ferrimagnetism and metallicity have been interpreted in terms of the indirect Fe(↑)-O-Mo(↓)-O-Fe(↑) exchange interactions [4], where localized (spin up) 3d electrons of $Fe^{3+}$ ions are mediated by the single itinerant (spin down) 4d electron of $Mo^{5+}$ ions. The electronic band structure calculations [5] also predicted a ferrimagnetic spin ordering with Fe ($3d^{5\uparrow}$) and Mo($4d^{\downarrow}$) states. But, the experimental value of saturated magnetization in $A_2FeMoO_6$ is always lower (~ 3.44, 3.52 and 3.84 in $\mu_B$/f.u for A = Ca, Sr and Ba, respectively) [3, 6] than the theoretical predicted value ($4\mu_B$/f.u.) for an ideal double perovskite structure with half-metallic state. Moreover, the experimental value is non-integer in comparison with the expected integer value ($4\mu_B$/f.u.), for perfect half-metallic state.

An attempt has been made to bridge the gap between experimental results and theoretical predicted value by introducing a finite amount of anti-site defects and anti-phase domains (APD) in the double perovskite structure [7-11, 4]. Such disorders have shown the effects to modify the ferrimagnetic Fe-O-Mo interactions or the indirect Fe-O-Mo-O-Fe exchange interactions and facilitate the formation antiferromagnetic Fe-O-Fe or Mo-O-Mo interactions at the antiphase boundary [12]. The effects of anti-site defect and anti-phase boundary can be understood by studying the structural, electrical and magnetic properties of $A_2FeMoO_6$ compounds with suitable non-magnetic substitution. D. Sanchez *et al* [13] reported a few interesting features by increasing La concentration (x) in the series of $Sr_{2-x}La_xFeMoO_6$. In addition to the usual decrease of magnetic moment, the $T_C$ of the material showed a non-monotonic increase as a function of increasing concentration (x) of La. The results were explained in terms of the decreasing site ordering of $Fe^{3+}$ and $Mo^{5+}$ ions (i.e., increase of anti-site defects) in $Sr_{2-x}La_xFeMoO_6$. Keeping in mind a rich variety of physical properties of La doped perovskite compounds $A_{1-x}La_xBO_3$ (A: Ca, Sr, Ba; B: Mn) [14-16], we propose to study a few more double perovskite systems for the understanding of La doping.

In the present work, we have studied the effect of La doping in $Ca_{2-x}La_xFeMoO_6$ compound. The special attention is to explore the structural aspects. We also briefly discuss the magnetism and electrical properties of selected samples and their dependence on the structural change in the material. The details of electrical and magnetic properties of the samples will be reported elsewhere.

## II. EXPERIMENTAL

The polycrystalline samples of $Ca_{2-x}La_xFeMoO_6$ series ($0.0 \leq x \leq 0.8$) have been prepared using the solid state sintering method. The stoichiometric amounts of high purity $La_2O_3$, $CaCO_3$, $Fe_2O_3$, $MoO_3$ and Mo powders were mixed and ground for 2 hours. The pellet form of the ground powder was sintered at different temperatures in the range - $900^0$C to $1200^0$C with intermediate slow cooling and heating. The crystal structure of the samples was characterized by x-ray diffraction spectrum at room temperature (300 K) using $CuK_\alpha$ radiation from X-ray diffractometer (model: Phillips-APD 1877). The lattice parameters and cell volume of the samples were obtained from cell refinement program (CELREF V3) and also cross checked by standard full profile fitting using FULLPROF Program. The scanning electron microscope (SEM) (model: HITACHI S-3400N, Japan) was employed to study the surface morphology of the samples. The elemental analysis of the samples was carried out using Energy Dispersive analysis of X-ray (EDX) spectrometer (Thermo electron corporation Instrument, USA). The magnetic properties of the samples were studied using SQUID magnetometer (MPMS7, Quantum Design, USA). The real and imaginary components of ac susceptibility at magnetic field $h_{rms}$=1 Oe and driving frequency ($\nu$) 1 Hz to 1000 Hz were measured in the temperature range 5 to 300 K. The field dependence of magnetization was measured at 5 K with applying field range ±70 kOe. The dc electrical resistivity of the samples was measured using the standard four probe technique in the temperature range 310 K to 900 K.

**III. Result and Discussion**

**A. Structural Properties**

Fig.1 shows the X-ray diffraction (XRD) pattern for various compositions of $La_xCa_{2-x}FeMoO_6$ compound. The XRD pattern of the samples appeared to be a single phase compound with double perovskite structure. A close inspection of the peaks (Fig. 2) at 2θ (in degrees) ~ 32.90 and 58.0 clearly indicated a minor splitting about the respective peak, which suggested the coexistence of a small fraction of secondary phase. The report by Song et al. [17] also indicated the coexistence of two phases, viz., orthorhombic phase and monoclinic phase, in $Ca_2FeMoO_6$ double perovskite structure. In their sol-gel derived $Ca_2FeMoO_6$ compound, the orthorhombic phase was reduced by increasing the annealing temperature from $900^0$C to $1100^0$C. In our material (Fig. 2), the wing like minor split (marked by dotted lines) about the main peak systematically decreases with La substitution and there is no split for x ≥ 0.5. The point is that all our samples were prepared in identical heat treatment. This result is interesting, because it indicates that the thermal effect is not the only factor which can reduce the minor secondary phases [17], but the increase of La substitution also shows the ability to suppress the secondary phase in $Ca_2FeMoO_6$ compound. The crystal structure in our solid state routed samples is dominated by the monoclinic structure with space group $P2_I$ /n. This result is significantly different in comparison with the sol-gel derived samples, but is consistent with other solid state routed sample [11,18]. This means the crystal structure depends up to certain extent on the preparation conditions. We have analyzed the XRD pattern of $La_xCa_{2-x}FeMoO_6$ samples by fitting the prominent peaks into a monoclinic crystal structure with space group $P2_I$/n. The calculated lattice parameters of the x = 0 ($Ca_2FeMoO_6$) sample (***a*** *~ 5.4154±0.0041,* ***b****~ 5.5219±0.0030,* ***c*** *~ 7.6894±0.0035* in Å unit and cell volume ***V*** ~ 228.46±0.47 Å$^3$) are well agreed with the reported results (***a*** *~5.4150* Å, ***b****~5.5224* Å, ***c****~7.7066* Å and ***V***~ 227.58 Å$^3$) [19]. The lattice parameters and cell volume (***a, b, c*** and ***V***) for different La substituted samples are shown in Fig. 3. We observe that ***a, c*** (presented as **c**/√2) and ***V*** increases with La substitution (x). At the same time, ***b*** slightly decreases with x. The over all increase of the lattice parameters and cell volume may be consistent with the fact that higher ionic radius $La^{3+}$ (1.16 Å) while replacing the smaller ionic radius $Ca^{2+}$ (ionic radius ~ 1.12 Å) increases the bond length in the crystal structure. Next we point out a systematic decrease of XRD peak intensity of the material with the increase of La substitution (x). The peak intensity appeared at 2θ (in degrees) ~19.86 (101) and ~ 38.86 (211) gradually lose their magnitude with La substitution and not traced out for x ≥ 0.5. The (312) peak (2θ ~$59.16^0$) intensity also gradually decreases and finally not seen for x ≥ 0.5. In $Sr_2FeMoO_6$ double perovskite structure these peaks were identified due to as super cell structure [20]. Similar structural properties, i.e., the increase of cell parameters and diminishing of certain XRD peaks, were reported for $Sr_{2-x}La_xFeMoO_6$ series [13]. The preliminary report on $Ca_{2-x}La_xFeMoO_6$ available in literature [21] also suggested similar structural change upon La doping. The fact was understood in terms of the increasing anti-site disorder (ASD) in double perovskite structure by La doping. For further structural information, we have analyzed the Scanning Electron Microscope (SEM) picture and Energy Dispersive Analysis of X-ray spectrum (EDX) at 300 K. We did not use any conductive coating on the surface of the used samples, because the samples showed the metallic properties. Fig. 4 (a-c) shows a significant change in the morphology of the material as the La concentration increases. The SEM picture in Fig. 4a shows that the particles are well compacted (melted) in the material to exhibit a relatively smooth (adhesive type) surface morphology for x = 0 sample. One could expect a minimum grain boundary disorder for this sample. The good quality SEM pictures in Fig. 4(b-c) suggest that particles in La doped samples are heterogeneously distributed in comparison with x = 0 sample brittle type. The particles are better distinct form each other and the surface roughness is more in La doped samples. A large grain boundary disorder is expected in La doped samples due to relatively lose contacted and non-uniform shaped particles in the material. Particle size of the samples appeared to be in the range of few µm. It would not be justified to provide an exact value of the particle size. However, the general trend of SEM picture suggested that a large fraction of smaller particles are formed in the La doped samples. Fig. 5 (a-f) shows one of the recorded EDX spectra and a line spectrum over a length of 50-55 μm, along with experimented sample zone, for x = 0 and 0.3 (representative of La doped) samples. The EDX spectrum identified Ca, Fe, Mo, O as constituent elements in the samples, in addition to La for doped samples. The obtained elemental composition is close to the expected value. The interesting result is that the atomic ratio of Mo/Fe is increasing with the increase of La doping (e.g., 0.77, 0.82, 0.89 for x = 0, 0.3 and 0.6 samples respectively). The feature also be confirmed by comparing the enhanced Mo peak at ~2.3 keV with respect to Ca peak ~3.5 keV in x = 0.3 sample in comparison with x = 0 sample. The Line spectrum of x = 0 sample (Fig. 5d) suggested a reasonably good chemical homogeneity within the statistical fluctuation of the counts due to uneven surface morphology of the sample. However, we noted a significant amount of exchange between the counts of Fe and Mo in the line position between 20 μm to 30 μm, which is beyond the statistical fluctuation of counts. This observation may be relevant to understand a significant amount of anti-site disorder (exchange of Fe and Mo between B and $B^{/}$ sublattices) in double perovskite structure prior to La substitution. On the other hand, the line spectrum for La, Fe and O is more or less uniform within statistical fluctuation in comparison with the spectrum of Ca and Mo for x = 0.3 sample (Fig. 5f). The statistical fluctuation in La doped samples is expected to be more due to large inhomogeneity in the surface morphology. However, the enormous loss of counts for all the elements at the line position near to 15 μm is attributed to the lack of sufficient counts to the detector from the deeper surface of the sample. Further discussion on the effects of La substitution is followed up using the magnetic and electric properties of the samples.

**B. Magnetic and Electric Properties**

The temperature dependence of real ($\chi^{/}$) and imaginary ($\chi^{//}$) components of ac susceptibility for selected samples, measured at ac field of 1 Oe and frequency range ν =1 Hz to 1 kHz, are shown in Fig.6 (a-d). The $\chi^{/}$ (T) of x = 0 sample increases with the increase of temperature from 5 K. The $\chi^{/}$ (T) exhibited a peak at about $T_P$ ~ 200 K and on further increase of temperature $\chi^{/}$ (T) rapidly decreases. On the other hand, $\chi^{//}$ (T) also showed similar behaviour with a peak at lower temperature $T_m$ ~100 K. The response of $\chi^{/}$ (T) to the variation of measurement frequency is weak, where as the frequency effect on the $\chi^{//}$ (T) data is well pronounced. The frequency response of $\chi^{/}$ (T) data is more prominent for x = 0.2 sample (Fig. 6b) in comparison with x = 0.4 (Fig. 6c) and 0.6 (Fig. 6d) samples. For x = 0.2 sample, the $\chi^{/}$ (T) data are well separated below 175 K with the variation of frequency, but the frequency dependence is less in the temperature range 175 K to 300 K. There is no peak behaviour for x = 0.2 sample up to 300 K and the peak in $\chi^{/}$ (T) may be expected at temperature higher than 300 K. In contrast, $\chi^{/}$ (T) data showed a well defined peak at about $T_P$ ~ 250 K and 220 K for x = 0.4 and 0.6 samples. Interestingly, $\chi^{//}$(T) for x = 0.2, 0.4 and 0.6 samples also exhibited a round shape peak near to the peak position ($T_m$ ~100 K) of x = 0 sample. The $T_m$ position is strongly frequency dependent for x = 0.2, 0.4 and 0.6 samples. In addition to increasing the magnitude of $\chi^{//}$ (T) peak at $T_m$, the peak position is shifted to higher temperature with the increase of frequency. At $\nu = 1$ Hz $T_m$ is ~110 K, 88 K, 123 K, 90 K and at ν = 1 κHz $T_m$ is ~120 K, 112 K, 148 K, 120 K for x = 0, 0.2, 0.4 and 0.6 samples, respectively. This gives the frequency shift per decade ($\Delta = \Delta T_m/[T_m(1\ Hz)\ \Delta ln\nu]$) ~ 0.013, 0.039, 0.029 and 0.047 for x = 0, 0.2, 0.4 and 0.6 samples, respectively. The obtained value of temperature shift per decade of frequency is larger than the typical value 0.001 for classical spin-glasses like CuMn and also less than the typical value 0.1 for ideal super paramagnetic system [22]. The obtained values of Δ are suggesting the coexistence of cluster glass type component in the samples. The coexistence of glassy component in x = 0 sample is interesting, because the sample was expected to be long range ordered ferrimagnet. The glassy feature of the La doped samples can be understood from the analysis of Vogel-Fulcher law: $\nu = \nu_0 exp[-E_m/(T_m-T_0)]$. Due to the lack of sufficient numbers of $T_m(\nu)$ points we do not intend for detailed analysis. However, we can roughly estimate the spin flip frequency ($\nu_0$) in the range $10^{10}$ Hz to $10^{14}$ Hz, activation energy ($E_m$) in the range 0.14 eV to 0.24 eV by varying the interaction term ($T_0$) in the range 20 K to 30 K. The estimated parameters suggest that the coexisting glassy component belongs in the cluster glass regime [23]. The temperature dependence of ac susceptibity data also provided all possible indications that paramagnetic to ferrimagnetic ordering temperature ($T_C$) is greater than room temperature (300 K) for all the samples and $T_C$ of the La doped samples appeared to be enhanced in comparison with x = 0 sample. These results are also consistent with some of the recent observations in identical compounds $Sr_{2-x}La_xFeMoO_6$ [13] and $Ca_{2-x}Nd_xFeMoO_6$ [24]. The next significant feature of La doping is that the magnitude of ac susceptibility ($\chi^{/}$ and $\chi^{//}$) of the material decreases with the increase of La concentration. This result is consistent with the experimental reports that non-magnetic La substitution has shown the effect of decreasing magnetic moment per formula unit in double perovskite structure. In fact, the field dependence of dc magnetization (Fig. 7) confirmed the decrease of magnetic moment with La doping. The field (H) dependence of magnetization (M) at 5 K for x = 0, 0.2, 0.4 and 0.6 samples indicated a typical soft (ferri) ferromagnetic nature, characterized by a sharp increase of M to achieve a saturation level above 10 kOe. The spontaneous magnetization ($M_S$) of the samples are estimated from the linear extrapolation of high field M(H) data to intersect on M axis at H = 0 value. The estimated spontaneous magnetization ($M_S$) ~ 3.25 $\mu_B$/f.u for x = 0 sample is well matched with the reported value (~3.44 $\mu_B$/f.u) at 4.2 K [6]. The spontaneous magnetization ($M_S$ ~ 3.25, 3.10, 2.75 and 1.30 in $\mu_B$/f.u for the x = 0.2, 0.4 and 0.6, respectively) decreases significantly at higher La doping. The M(H) data also exhibited hysteresis loop at 5 K and a typical portion of the loop is shown in the inset of Fig. 7 for x = 0, 0.2 and 0.4 samples. We noted a typical value of remanent magnetization $M_R$ ~0.16, 0.24, 0.21 and 0.14 $\mu_B$/ f.u and coercivity $H_C$ ~32, 78, 79 and 86 Oe for x = 0.0, 0.2, 0.4 and 0.6 samples, respectively. The decrease of $M_R$ is consistent with the decrease of magnetic moment in La doped samples. The interesting feature is that magnetic hardness, represented by $H_C$, of the material also increases with the increase of La doping. This result is consistent with the increasing disorder in the material due to La doping. The increasing grain boundary disorder with the increase of La substitution, as suggested from the SEM results, is also confirmed from the increasing resistivity of the material in the temperature range 310 K to 900 K. The temperature dependence of resistivity ($\rho$) data are normalized by 310 K data and the plots are shown in Fig. 8. The slope ($\alpha$) of the $\rho$(T) data was calculated from linear fit of $\rho$(T) data above 600 K, where the samples are expected to be in paramagnetic state and the resistivity of a sample is independent of magnetic spin ordering. In this temperature regime, the resistivity is strongly influenced by other effects, such as: grin boundary contributions and internal defects introduced by anti-site disorder and structural distortion, in highly conductive double perovskite materials. The $\rho$(310 K) and slope ($\alpha$) for different samples are shown in the inset of Fig. 8. The positive slope ($\alpha$) for all the samples indicated the retaining of metallic behaviour even in the La doped samples. However, the resistivity at 310 K $\rho$(310 K) increases [e.g., $\rho$(310 K) ~0.639 and 2.512 in $\Omega$-cm for x = 0 and 0.6 respectively] and the calculated slope showed decreasing trend (e.g., $\alpha$ ~ 10.35 and 0.94 in$10^{-3}$ $K^{-1}$ for x = 0 and 0.6, respectively) with the increase of La concentration. This gives further evidence of increasing disorder in $Ca_2FeMoO_6$ structure as an effect of La doping. The effect of increasing disorder has been relayed to the low temperature (< 300 K) state and already reflected in the increasing temperature shift per decade of frequency value ($\Delta$) for La doped samples. In fact, the $\Delta$ value of x = 0.4 sample is close to the insulating spin-glass system (EuSr)S ~ 0.06 [22]. The breaking of large size ferromagnetic grains into smaller grains (in terms of different size clusters) was indicated from SEM picture and this may cause a significant increase of resistance in spite of metallic character, in the doped samples.

## IV. SUMMARY AND CONCLUSIONS

The present experimental work deals with the samples of La doping in $Ca_2FeMoO_6$. The samples showed stability within the double perovskite structure. Some internal structural change (e.g., suppression of specific XRD peaks, expansion of cell volume, accommodation of more Mo atoms in the crystal structure, loss of surface smoothness, enhanced grain boundary contribution by increasing the number of grains) was noted with the increase of La concentration. These structural changes played a significant role in modifying the magnetic and electrical properties of the material. In literature these effects have been taken care in terms of anti-site disorder and anti-phase domains (APD) in the double perovskite structure. We believe that these structural defects (with respect to ideal double perovskite structure) are always present in the parent $A_2FeMoO_6$ compound, which is independent of material synthesis. The secondary phase is inevitable both in chemical and solid state routed samples. For chemical routed samples, the major contribution to secondary phase coming from a significant fraction of precipitation which is not well melted in the main structure of the compound or due to lack of complete diffusion of elements at lower temperatures. On the other hand, in solid state route sample, the secondary component is contributed by the loss of a fraction of certain element in the compound, e.g., Mo in our samples. The introduction of intrinsic disorder in double perovskite structure is also highly possible from a finite amount of lattice strain generated to accommodate two perovskite units ($AFeO_3$ and $AMoO_3$) in rock-salt (NaCl) type arrangements. The second option may be less affected by the synthesis route. Anyway, the magnetic moment of the material decreases with the increase of structural defects and super cell contributions is compensated in the double perovskite structure. The super cell contribution may arise due to the significant loss of Mo atoms in the structure and the loss is minimized in the expanded cell volume as an effect of La doping. The increase of La concentration also showed similar results in $Sr_{2-x}La_xFeMoO_6$ [13]. It seems that the effects of La substitution in the A sublattice of $A_2FeMoO_6$ double perovskite structure may be of universal nature. This conclusion needs more experimental supports. Some other aspects upon La doping, such as modified exchange interactions [18] and band filling effect [24], is not discussed in this work. We feel that such discussions are more appropriate by correlating the low temperature magnetic and electrical properties, which we have already studied and to present in future article.

In conclusion, we noted some interesting structural and physical properties in $Ca_{2-x}La_xFeMoO_6$**.** The physical (i.e., magnetic and electrical) properties are strongly correlated with the internal change of crystal structure and morphology. We attribute the appearance of cluster glass component in the ferrimagnetic spin structure and increasing resistivity in the metallic network as an important consequence of increasing disorder in the double perovskite structure.

ACKNOWLEDGMENT

We thank Central Instrument Facility, Pondicherry University for providing some of the material characterization facility.